\documentclass[journal=jacsat,manuscript=article]{achemso}
\usepackage[version=3]{mhchem} 
\usepackage{graphicx,dcolumn,bm,siunitx,float,amssymb}
\usepackage{makecell,multirow,xcolor,ifthen}
\usepackage[normalem]{ulem}

\usepackage[pdfpagelabels, pdfencoding=auto, psdextra]{hyperref}
\hypersetup{pdfsubject=Paper, pdfkeywords={nuclear physics} 
 unicode = true, breaklinks = true, colorlinks = true,
 linkcolor = blue, menucolor = blue, citecolor = blue,
 urlcolor = blue}

\newboolean{showred}
\newboolean{showblue}
\newboolean{showdelete}
\setboolean{showred}{false}   
\setboolean{showblue}{false}   
\setboolean{showdelete}{false}

\newcommand{\highlightred}[1]{%
  \ifthenelse{\boolean{showred}}{\textcolor{red}{#1}}{#1}}
\newcommand{\highlightblue}[1]{%
  \ifthenelse{\boolean{showblue}}{\textcolor{blue}{#1}}{#1}}
\newcommand{\deletetext}[1]{%
  \ifthenelse{\boolean{showdelete}}{\sout{#1}}{}}

\author{Dip Joti Paul}
\affiliation{Research Laboratory of Electronics, Massachusetts Institute of Technology, Cambridge, MA 02139, USA}
\email{djpaul@mit.edu}

\author{Stewart Koppell}
\affiliation{Research Laboratory of Electronics, Massachusetts Institute of Technology, Cambridge, MA 02139, USA}

\author{Gregor G. Taylor}
\affiliation{Jet Propulsion Laboratory, California Institute of Technology, Pasadena, CA 91109, USA}
\altaffiliation{Advanced Quantum Architecture Laboratory (AQUA), École polytechnique fédérale de Lausanne (EPFL), Neuchâtel 2002, Switzerland}

\author{Boris Korzh}
\affiliation{Jet Propulsion Laboratory, California Institute of Technology, Pasadena, CA 91109, USA}

\author{Sahil R. Patel}
\affiliation{Jet Propulsion Laboratory, California Institute of Technology, Pasadena, CA 91109, USA}

\author{Andrew D. Beyer}
\affiliation{Jet Propulsion Laboratory, California Institute of Technology, Pasadena, CA 91109, USA}

\author{Emma E. Wollman}
\affiliation{Jet Propulsion Laboratory, California Institute of Technology, Pasadena, CA 91109, USA}

\author{Matthew D. Shaw}
\affiliation{Jet Propulsion Laboratory, California Institute of Technology, Pasadena, CA 91109, USA}

\author{Phillip D. Keathley}
\affiliation{Research Laboratory of Electronics, Massachusetts Institute of Technology, Cambridge, MA 02139, USA}

\author{Karl K. Berggren}
\affiliation{Research Laboratory of Electronics, Massachusetts Institute of Technology, Cambridge, MA 02139, USA}
\email{berggren@mit.edu}

\title[]{Enhanced Mid-Infrared Single-Photon Detection with Antenna-Coupled Superconducting Nanowires}

\begin{document}
\begin{abstract}
Scaling the photon-detection area of superconducting nanowire single-photon detectors (SNSPDs) has traditionally been achieved by nanowire meandering. However, material inhomogeneities and fabrication-induced defects, such as line-edge roughness, increase with nanowire length, leading to reduced internal photon-detection efficiency and elevated dark-count rates. This trade-off becomes increasingly pronounced as nanowires are scaled to sub-100 nm widths and sub-5 nm thicknesses required for mid- to far-infrared sensitivity. Here, we demonstrate an antenna-coupled SNSPD architecture that enhances the effective photon-detection area without increasing nanowire length. A crossed bowtie antenna integrated with an 80 nm-wide, 3 nm-thick WSi nanowire yields 15.7× increase in effective detection area at $\SI{7.4}{\micro\meter}$ compared to a bare nanowire of identical geometric footprint, while maintaining the same internal detection efficiency and dark-count rate. Antenna coupling provides a scalable approach to increasing photon-detection area while reducing the noise-equivalent power, offering performance benefits for applications in astronomy, biological imaging, and molecular spectroscopy.
\end{abstract}


\vspace{-0.3cm}
\section{Introduction}
\vspace{-0.3cm}
Mid-infrared (mid-IR) imaging has enabled a wide range of scientific and industrial applications in recent decades by facilitating the thermal mapping of objects and the identification of biomolecular and chemical compounds \cite{lau2023superconducting}. Consequently, high-performance mid-IR detectors are increasingly in demand across multidisciplinary fields \cite{mccarthy2025high, wollman2024current, koppell2026dark}. Superconducting nanowire-based detectors have demonstrated excellent performance in sensitive, time-resolved single-photon detection in the near-infrared spectrum \cite{marsili2012efficient, mccarthy2025high, korzh2020demonstration}. However, they are not yet a mature platform for mid-IR and far-IR wavelengths ($> \SI{5}{\micro\meter}$). Recent demonstrations of spectral sensitivity in superconducting nanowire single-photon detectors (SNSPDs) at \SI{29}{\micro\meter} wavelength have attracted significant attention \cite{taylor2023low, hampel2025tungsten}, positioning this technology as a promising path for mid-infrared single-photon detection among competing superconductor-based detectors \cite{lau2023superconducting, taylor2026mid}. However, key metrics such as photon-absorption efficiency, photon-detection area, and noise-equivalent power still require improvement.

The prevailing approach to extending SNSPD sensitivity to mid-IR photons has been to decrease the critical temperature of the superconducting films by modifying the film stoichiometry \cite{verma2021single,taylor2023low}, which comes at the cost of operating the detector at lower temperatures, typically below 1 K. In parallel, superconducting nanowires are fabricated to be narrower (sub-100 nm) and thinner (sub-5 nm) to enhance mid-IR sensitivity \cite{marsili2012efficient,colangelo2022large,taylor2023low}. This dimensional scaling, however, reduces the nanowire’s photon-absorption efficiency and increases its susceptibility to nanofabrication imperfections. As a result, large-area meanders often fail to achieve saturated internal detection efficiency at mid-IR \cite{marsili2012efficient,colangelo2022large,taylor2023low}. Compared to short nanowires, long nanowire meanders are more susceptible to material inhomogeneities and nanofabrication-induced linewidth variations \cite{liu2025revealing}, often referred to as constrictions or weak points of the detector, which can significantly reduce the bias margin and impede achieving saturated detection efficiency \cite{colangelo2022large,taylor2023low}. In addition, bends in meandered nanowires can act as potential sources of constriction \cite{clem2011geometry}, and meandered nanowires have been observed to exhibit higher dark-count rates than short nanowires biased at similar fractions of their switching currents \cite{colangelo2022large,taylor2023low}.

In this work, we demonstrate that resonant metallic antennas on photon-sensitive superconducting nanowires offer a promising alternative to the conventional nanowire meander for scaling the photon-detection area at mid-IR. In this antenna-coupled configuration, the antennas capture and concentrate incoming free-space radiation within their feed gaps, thereby significantly increasing the effective photon-detection area of the superconducting nanowire beyond its geometric area. Thus, the antennas serve as photon collectors, while the nanowires function solely as photon sensors, unlike conventional meandered geometries where the nanowire serves both roles. Additionally, conventional SNSPD meanders exhibit intrinsic sensitivity to the polarization of the incident radiation, which can be engineered through antenna coupling. Antennas can function as spectral filters, and by leveraging this capability, arrays of tailored antennas coupled to photon-sensitive nanowires can enable polarization-insensitive detection with spectral and incident-angle selectivity.

Enhancing the photon-absorption efficiency of SNSPDs through antenna coupling has previously been demonstrated at 1550 nm, yielding 1.5× to 2.3× improvements in photon-detection efficiency \cite{hu2010superconducting, heath2015nanoantenna}. Our work extends this concept to the mid-IR, showing 15.7× increase in detection efficiency at \SI{7.4}{\micro\meter} through antenna coupling. Conventional cavity-based absorption-enhancement schemes (e.g., distributed Bragg reflectors) become increasingly challenging at mid- and far-IR wavelengths \cite{deng2026silicon} because of absorption losses in the constituent materials, increased cavity thickness, and limited refractive-index contrast, thereby providing strong motivation for adopting the antenna-coupling approach in this regime. We fabricated and characterized antenna-coupled WSi nanowires and observed 15.7× higher photon-absorption efficiency than in an otherwise identical WSi nanowire without an antenna at 7.4 µm, while maintaining similar dark-count rates. We then performed simulations to investigate the scalability of this approach for larger detection areas based on periodic antenna arrays. Our analysis indicates that antenna-coupled nanowires in a periodic array configuration can exhibit significantly improved detection sensitivity, with an approximately 900× reduction in noise-equivalent power (NEP) at 7 µm wavelength relative to a conventional meandered nanowire of identical 50 × 50 µ\si{m^2} geometric footprint \highlightblue{under well-shielded measurement conditions.}


\vspace{-0.3cm}
\section{Results and Discussion}
\begin{figure}[!htbp]
    \vspace{-0.5cm}
    \centering
    \includegraphics[width=0.95\textwidth]{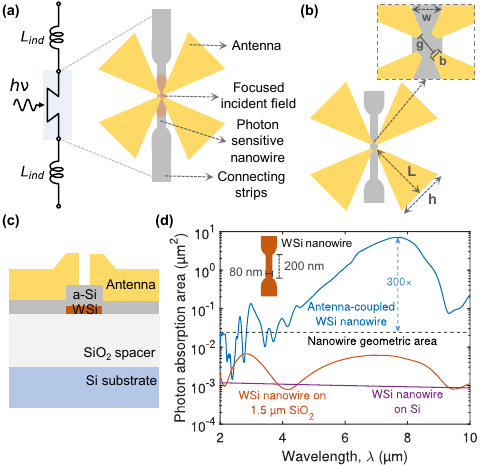}
    \vspace{-0.2cm}
    \caption{\small \textbf{(a)} A schematic of an antenna-coupled SNSPD is shown, featuring a narrow superconducting constriction in series with inductors ($L_{ind}$) to enable self-reset operation of the device. A geometric view of the device depicts a crossed-bowtie antenna patterned on top of a superconducting nanowire that serves as the photon-sensitive region, with the antenna acting as the photon collector. \textbf{(b)} Top-view of a crossed-bowtie antenna is shown, where $L$ denotes the antenna arm length, $h$ the arm width, $b$ the apex width, $g$ the antenna feed gap, and $w$ the nanowire width. \textbf{(c)} Vertical cross-section of an antenna-coupled nanowire is shown, with \si{SiO_2} acting as a dielectric spacer on the silicon substrate, and amorphous silicon serving as an encapsulation layer to prevent the electrical and thermal contact between the WSi superconducting nanowire and the gold antenna. \textbf{(d)} The effective photon-absorption area of a 3 nm-thick, 80 nm-wide, and 200 nm-long WSi nanowire as a function of wavelength is simulated for three different device configurations. The dashed line represents the nanowire's geometric area (80 nm × 200 nm), while the solid colored lines show the nanowire’s effective photon-absorption area for each configuration. Antenna coupling can significantly increase the nanowire's effective photon-absorption area beyond its geometric footprint (by nearly 300× at the antenna resonant wavelength), thereby enabling the scaling of the nanowire’s effective photon-detection area without increasing its physical footprint. \vspace{-3mm}}
    \label{fig:Fig1}
\end{figure}

\vspace{-0.3cm}
Figure \ref{fig:Fig1}a presents a schematic of an antenna-coupled superconducting nanowire, in which a crossed bowtie antenna is placed on top of the superconducting nanowire, with its narrow feed gap centered on the nanowire to maximize coupling to the incident free-space radiation. As the photon-sensitive nanowire segment can be very short (i.e., sub-micron in length), the inclusion of series inductors ($L_{ind}$) may be necessary to increase the inductance of the detector and prevent latching after photon-detection events. In this work, we employ a crossed bowtie antenna due to its simple geometry and relatively broadband resonance. Figure \ref{fig:Fig1}b shows metallic (i.e., gold) bowtie antennas on a photon-sensitive superconducting nanowire, where $L$ denotes the antenna arm length, $h$ the arm width, $b$ the apex width, and $g$ the antenna feed gap. The feed gap is typically chosen to be smaller than the nanowire width to maximize overlap between the localized field and the photon-sensitive nanowire.

A vertical cross-section of the antenna-coupled WSi superconducting nanowire is shown in Figure \ref{fig:Fig1}c. Since metallic antennas are good electrical and thermal conductors, a thin insulating layer is required between the antenna and the superconducting nanowire to provide electrical and thermal isolation. Without this layer, the metallic antennas would act as a low-resistance shunt for the superconducting nanowire, complicating the reset and readout of the detector. Materials with low mid-IR absorption, such as amorphous silicon (a-Si), \si{CaF_2}, and \si{MgF_2}, are therefore well suited for this insulating layer, allowing photons collected by the antenna to be absorbed primarily in the superconducting nanowire. In this work, a-Si is used due to its relatively high refractive index and low mid-IR absorption, which improves optical mode confinement and enhances the field overlap between the antenna and the nanowire. To enhance the radiative characteristics of the antennas, a quarter-wavelength-thick dielectric spacer (i.e., \si{SiO_2}) beneath the antenna-coupled nanowire can be considered \cite{seok2011radiation,brown2015fan}, as shown in Figure \ref{fig:Fig1}c. Since bowtie antennas function as electric dipoles, the oxide spacer can improve the local field enhancement by enabling constructive interference between the antenna-scattered fields and the reflections from the silicon substrate. 

Figure \ref{fig:Fig1}d shows the photon-absorption area, or effective aperture, of a short superconducting nanowire for different device configurations, calculated using finite-difference time-domain (FDTD) simulations. In all cases, the superconducting nanowire remains the same, consisting of an 80 nm-wide, 200 nm-long, 3 nm-thick WSi nanowire encapsulated by a 7 nm-thick a-Si layer. The photon-absorption area was determined by calculating the ratio of the optical power absorbed by the nanowire to the illumination intensity of the incident optical field. Three configurations were simulated: (i) a resonant antenna-coupled WSi nanowire on a \SI{1.5}{\micro\meter}-thick \si{SiO_2} spacer on a silicon substrate (blue line in Figure \ref{fig:Fig1}d), (ii) a WSi nanowire on the same \si{SiO_2}/silicon substrate without antenna coupling (red line), and (iii) a WSi nanowire on a silicon substrate without antenna coupling (violet line). In Figure \ref{fig:Fig1}d, the dashed line indicates the nanowire's geometric area of 80 nm × 200 nm, so a photon-absorption area equal to this value corresponds to 100\% absorption of photons incident on the nanowire’s geometric area.

For the bare WSi nanowire on a silicon substrate, the effective photon-absorption area decreases monotonically with wavelength, corresponding to the monotonic decrease in the nanowire’s photon-absorption efficiency in the mid-IR. Introducing a \SI{1.5}{\micro\meter}-thick \si{SiO_2} spacer without antenna coupling leads to enhanced photon absorption in WSi nanowire at specific wavelengths due to the resonant cavity effects; however, even at resonance, the absorption area does not exceed the nanowire’s geometric area, showing that the effective photon-detection area of the nanowire remains constrained by its physical footprint.

In contrast, for the antenna-coupled nanowire configuration, the nanowire’s effective photon-absorption area exceeds its geometric area for wavelengths greater than \SI{4}{\micro\meter} and reaches nearly $300\times$ the nanowire’s geometric area at \SI{7.7}{\micro\meter} wavelength. Unlike the planar dielectric cavities, which rely on standing-wave field enhancement and therefore cannot achieve photon-absorption area exceeding the absorber footprint, antenna coupling can exceed this limit through near-field collection and concentration. The effective absorption area enhancement can be further increased by incorporating a metallic (i.e., Au) reflector beneath the \si{SiO_2} spacer, which enables strong constructive interference of the antenna-scattered fields by inducing an image dipole of opposite charge in the Au reflector \cite{seok2011radiation}. Overall, these simulation results show that resonant metallic antennas can act as effective photon collectors for photon-sensitive nanowires at mid-IR and enable substantial scaling of the detection area while keeping the nanowire geometric footprint small.

To demonstrate the mid-IR photon-detection area scaling through antenna coupling, we fabricated antenna-coupled SNSPDs on a 3 nm-thick WSi thin film deposited on a \SI{1.5}{\micro\meter} thermal oxide-coated silicon substrate. The film stoichiometry was 50:50 W:Si and was sputtered at 130 W and 5 mTorr. A 3 nm a-Si layer was deposited in situ as a passivation layer. The film had a sheet resistance of 850 $\Omega/\square$ at room temperature and a critical temperature (\si{T_c}) of 2.6 K.

\begin{figure}[!htbp]
    \vspace{-0.5cm}
    \centering
    \includegraphics[width=0.95\textwidth]{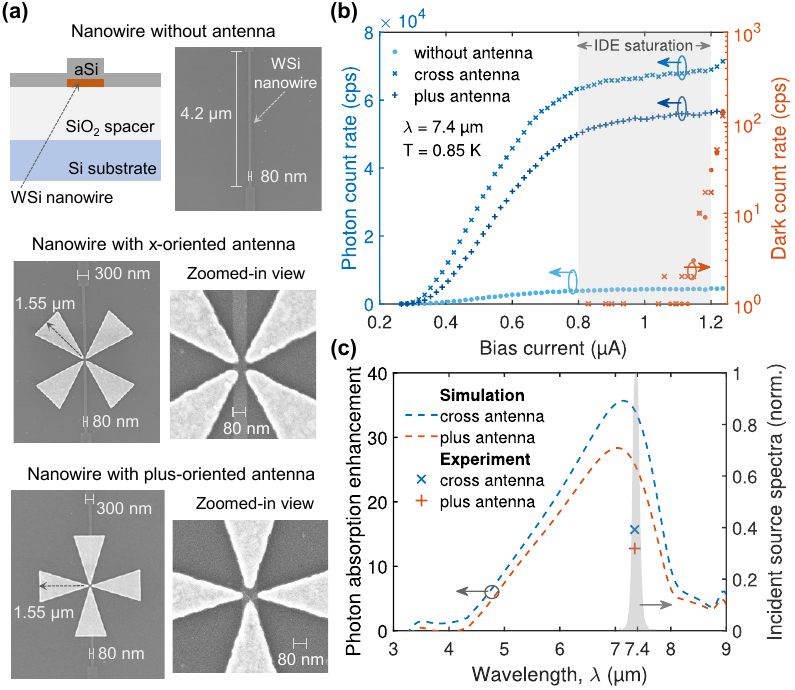}
    \vspace{-0.2cm}
    \caption{\small Measurements of the photon-count rate (PCR) and dark-count rate (DCR) for two configurations of antenna-coupled nanowires, compared with a nanowire of identical dimensions without an antenna on the same chip, under the same illumination and measurement conditions. \textbf{(a)} Vertical cross-section of the material stack and a scanning electron microscopy (SEM) image of an 80 nm-wide, 4.2 µm-long bare nanowire are shown along with SEM images of antenna-coupled nanowires of the same dimensions with cross- and plus-oriented antennas. In both configurations, the bowtie antenna has an arm length of \SI{1.55}{\micro\meter}. The zoomed-in images show that the antenna feed gaps in both of the fabricated antennas are slightly offset from the center of the nanowire width. \textbf{(b)} The PCR measurements show about 15.7× and 12.8× higher count rates for the cross- and plus-oriented antenna-coupled nanowires, respectively, compared to the bare nanowire at the same bias current, while their nearly identical internal detection efficiency (IDE) saturation plateau and DCRs indicate that the photon-sensitive nanowire segments are effectively the same in all three devices. \textbf{(c)} FDTD-simulated photon-absorption enhancement of the nanowire due to antenna coupling is plotted as a function of wavelength (dashed lines), while the cross and plus symbols represent the enhancement values for the cross- and plus-oriented antennas extracted from the saturated regions of their PCR data, normalized to the saturated PCR data of the bare nanowire. The normalized spectrum of the filtered thermal source incident on the sample is shown as the shaded grey region, centered at \SI{7.4}{\micro\meter}. \vspace{-3mm}}
    \label{fig:Fig2}
\end{figure}

Figure \ref{fig:Fig2}a shows the sample stack and scanning electron microscopy (SEM) images of three devices: an 80 nm-wide, \SI{4.2}{\micro\meter}-long bare WSi nanowire, an identical nanowire with cross-oriented bowtie antennas, and one with plus-oriented antennas. The connecting strips were 300 nm wide, nearly four times the nanowire width, so only the 80 nm wide segment contributed to photon counts. All three devices were fabricated within a 1 mm × 1 mm area of the chip and were sufficiently close to experience similar illumination intensity. The fourfold-symmetric crossed bowtie antenna is polarization-insensitive and optimal for unpolarized light. The antenna arm length is \SI{1.55}{\micro\meter}, width \SI{1}{\micro\meter}, apex width 20 nm, and feed gap 40 nm. The antennas were fabricated using a metal lift-off process with a 45 nm-thick gold layer and a 2 nm-thick titanium adhesion layer. Detailed fabrication steps are provided in Section V of the Supporting Information.

To characterize mid-IR single-photon sensitivity, we used a broadband thermal source and a stack of optical filters, with details provided in Section VI of the Supporting Information. Based on the room-temperature filter characteristics, the filtered spectrum incident on the devices was expected to be centered at \SI{7.4}{\micro\meter} with 148 nm full-width at half-maximum (FWHM), shown by the shaded region in Figure \ref{fig:Fig2}c.

Figure \ref{fig:Fig2}b compares the photon-count rate (PCR) and dark-count rate (DCR) of antenna-coupled and bare nanowires under identical illumination and operating temperature. All three devices exhibited a long PCR plateau, indicating saturated internal detection efficiency. In this regime, differences in PCR directly reflect relative photon-absorption efficiency. Hence, the enhancement due to antenna coupling can be determined from the ratio of PCR values in the saturation plateau (for bias currents from \SI{1}{\micro\ampere}--\SI{1.2}{\micro\ampere}) of the antenna-coupled nanowires to that of the bare nanowire. The average PCR ratios were 15.7× for the cross-oriented antenna and 12.8× for the plus-oriented antenna, indicating corresponding increases in effective photon-detection area relative to the identical-dimension bare nanowire.

As shown in Figure \ref{fig:Fig2}b, the DCRs of all three devices were nearly identical. Since the intrinsic DCR primarily arises from material inhomogeneity and nanofabrication-induced linewidth variations \cite{liu2025revealing,zhang2022geometric,andreev2024dark}, the similar intrinsic DCR values indicate comparable nanowire quality across the antenna-coupled and bare nanowire devices. Thus, resonant antenna coupling can significantly increase photon-absorption efficiency without degrading nanowire homogeneity. \highlightblue{Notably, antenna coupling can enhance the absorption of background photons within its resonance bandwidth. However, we did not observe any measurable increase in the background count rate, suggesting that resonant background photons were negligible under our measurement conditions.} 

Figure \ref{fig:Fig2}c compares the experimental photon-absorption enhancement with FDTD simulations. At \SI{7.4}{\micro\meter}, the simulated enhancement factors are about 33× for the cross-oriented antenna and 25× for the plus-oriented antenna, exceeding the measured values of 15.7× and 12.8×, respectively. The higher enhancement of the cross-oriented antenna likely arises from stronger spatial overlap between the antenna near-field and the WSi nanowire than that in the plus-oriented antenna. Notably, the measured enhancement factors for both devices are lower than the simulated values by approximately the same factor, suggesting that the discrepancy arises from effects common to both structures. Possible causes include fabrication imperfections, such as feed-gap misalignment with the nanowire center (as shown in the zoomed-in view in Figure \ref{fig:Fig2}a), surface contamination at the antenna–nanowire interface, or \highlightblue{a thicker-than-designed a-Si layer between the antenna and nanowire that may reduce the antenna near-field overlap with the WSi nanowire and increase absorption loss in the a-Si}. In addition, the spectral response of the bandpass filter, characterized at room temperature, may shift to longer wavelengths at cryogenic temperatures and contribute to the mismatch. The relative contributions of these effects, however, require further investigation.

It is worth noting that the observed absorption enhancement depends strongly on the length of the antenna-coupled nanowire. For the fabricated \SI{4.2}{\micro\meter}-long, 80 nm-wide WSi nanowire coupled to a cross-oriented bowtie antenna, the simulated enhancement at resonance is approximately 33× relative to an identical nanowire without an antenna (Figure \ref{fig:Fig2}c), whereas the much shorter 200 nm-long, 80 nm-wide nanowire in Figure \ref{fig:Fig1}d shows an enhancement of $\sim$300× relative to its geometric area and $\sim$1200× relative to an identical nanowire without an antenna. This comparison indicates that the antenna-induced field enhancement is highly localized near the feed gap, so only a short nanowire segment interacts strongly with the antenna near-field. Accordingly, the photon-sensitive nanowire length should be comparable to this antenna-enhanced region, since extending it further provides little additional absorption and may increase dark counts. Further details are provided in Section III of the Supporting Information.

\begin{figure}[!htbp]
    \vspace{-0.5cm}
    \centering
    \includegraphics[width=\textwidth]{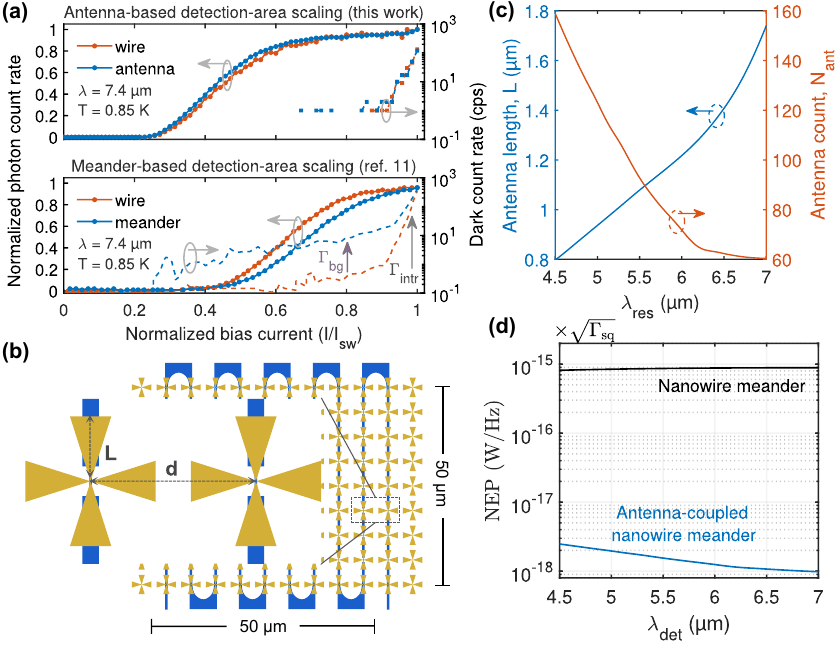}
    \vspace{-0.7cm}
    \caption{\small \textbf{(a)} Normalized PCR and unnormalized DCR of the 80 nm-wide, 4.2 µm-long WSi nanowires reported in this work, one antenna-coupled and one bare (without an antenna), are shown. For comparison, PCR and DCR data of WSi-based SNSPDs reported by Colangelo \textit{et al.} \cite{colangelo2022large}, featuring a 60 nm-wide, 20 µm-long nanowire and a 60 nm-wide, 560 µm-long meander, are shown (Adapted from Ref. \cite{colangelo2022large}; data were extracted and replotted from that work. Copyright 2022 American Chemical Society). These results show that, compared with a bare nanowire, antenna coupling increases the effective photon-detection area without changing the bias-dependent PCR or DCR behavior, whereas increasing the detection area by nanowire meandering substantially increases the background-dominated dark-count rate ($\Gamma_\mathrm{bg}$) and also affects the PCR saturation plateau. \textbf{(b)} Schematic of a periodic antenna-coupled nanowire meander spanning 50 × 50 µ\si{m^2} area is shown, where the antenna-coupled photon-sensitive nanowires are interconnected by wide, non-photon-sensitive tapers. \protect\highlightblue{The antenna array was extended beyond the photon-sensitive nanowire region to more closely approximate an infinite periodic array.} The inset shows two antenna-coupled nanowire segments from the array; the antenna length is denoted by $L$ and the array pitch by $d$. \textbf{(c)} Antenna length ($L$) required for resonance at $\lambda_\mathrm{res}$ in a periodic antenna-coupled nanowire meander geometry is shown. The number of antennas ($N_{\mathrm{ant}}$) required to span 50 × 50 µ\si{m^2} geometric footprint at the optimal array pitch is plotted, showing that fewer antennas are needed to span the same geometric footprint as the antenna length and pitch increase with wavelength. \textbf{(d)} Calculated noise-equivalent power (NEP) as a function of the SNSPD detection wavelength ($\lambda_\mathrm{det}$) for a conventional 80 nm-wide nanowire meander (25\% filling factor) and a periodic antenna-coupled nanowire meander comprising 80 nm-wide, 160 nm-long nanowire segments at the optimal array pitch, both spanning a 50 × 50 µ\si{m^2} geometric footprint. Here, $\Gamma_{\mathrm{sq}}$ denotes the dark-count rate per unit square length of the nanowire (Hz/$\square$). \vspace{-3mm}}
    \label{fig:Fig4}
\end{figure}

Next, we examine how antenna coupling provides an effective route to scaling the photon-detection area in the mid-IR compared to conventional meander geometry. The top panel of Figure \ref{fig:Fig4}a shows the normalized PCR and unnormalized DCR from Figure \ref{fig:Fig2}b for the plus-oriented antenna-coupled nanowire and the bare nanowire, showing that antenna coupling increases the effective photon-detection area without changing the bias-dependent PCR or DCR behavior.   

The bottom panel of Figure \ref{fig:Fig4}a shows the normalized PCR and unnormalized DCR of a 60 nm-wide, 20 µm-long straight wire and a 60 nm-wide, 560 µm-long meandered wire reported by Colangelo \textit{et al.} \cite{colangelo2022large}. Despite the fabrication process being optimized to achieve a low line-edge roughness of \(\sim\)4.9 nm in the meandered wire \cite{colangelo2022large}, the PCR saturation plateau shifts to higher bias currents in the meandered device compared to the 20 µm-long straight wire. This shift indicates reduced internal photon-detection efficiency in the meandered wire relative to the straight wire, likely due to accumulated nanowire nonuniformity arising from line-edge roughness, material inhomogeneities, and bend-induced current crowding \cite{liu2025revealing,clem2011geometry}. The unnormalized PCR was nearly 28× higher in the meandered configuration than in the straight nanowire, indicating that the photon-detection area scales with nanowire length \cite{colangelo2022large}. 

However, the DCR was also higher in the meandered device than in the straight nanowire under the same measurement conditions and normalized bias current \cite{colangelo2022large}. In Figure \ref{fig:Fig4}a, two distinct DCR regimes are observed. At lower normalized biases, the DCR shows weak bias dependence and is dominated by background-induced dark counts ($\Gamma_\mathrm{bg}$) arising from stray infrared photons. In this regime, the DCR depends on experimental factors such as radiation shielding in the cryogenic setup and device packaging and can be reduced by implementing cryogenic narrowband filtering and improved device packaging \cite{taylor2023low}. Within the same experimental conditions, $\Gamma_\mathrm{bg}$ scales approximately with the total nanowire length, as evidenced by the nearly 28× increase in background-induced DCR in the meandered device compared to the straight nanowire \cite{colangelo2022large}. In contrast, the antenna-coupled device shows no \highlightblue{measurable} increase in $\Gamma_\mathrm{bg}$ relative to the bare nanowire, \highlightblue{suggesting that background photons within the antenna resonance bandwidth were negligible under our measurement conditions. However, further investigation of background-photon coupling in antenna-coupled SNSPDs remains an important direction for future work.}

As the normalized bias current approaches unity, the DCR enters a steep exponential regime dominated by intrinsic vortex-related mechanisms, such as thermally activated vortex entry at the most current-crowded or defect-susceptible regions of the device \cite{liu2025revealing, zhang2022geometric,andreev2024dark}. In this regime, the intrinsic dark-count rate ($\Gamma_\mathrm{intr}$) converges to nearly the same exponential slope for both meandered and straight nanowires, indicating that one or a few weak spots, rather than the total device length or geometry, set the dark-count behavior at high bias. Hence, a similar exponential dependence is also observed for the antenna-coupled and bare devices. 
 
We next examine the scaling of an antenna-coupled nanowire to a large detection area and compare its detection sensitivity with that of a conventional nanowire meander of identical geometric footprint. Although antenna coupling increases the effective photon-detection area of a short nanowire, scaling this configuration to large geometric footprints requires integrating multiple antenna-coupled units in series. Figure \ref{fig:Fig4}b shows such an architecture: a periodic antenna-coupled nanowire meander spanning 50 × 50 µ\si{m^2}, where each photon-sensitive nanowire segment is coupled to a plus-oriented bowtie antenna, and adjacent segments are connected by wide non-photon-sensitive tapers. The inset shows two neighboring antenna-coupled nanowires separated by the antenna-array pitch ($d$). 

In this periodic antenna-coupled nanowire configuration, the array pitch sets the spacing between neighboring photon-sensitive nanowires. Closely spaced antennas enhance absorption through collective near-field interactions, but high antenna density increases the total photon-sensitive nanowire length, making the device more susceptible to background-induced DCR and line-edge roughness. Conversely, increasing the pitch reduces antenna density and improves scalability but eventually lowers absorption as antennas behave as isolated scatterers. Therefore, an optimal pitch should be evaluated at each antenna resonant wavelength to balance photon-absorption efficiency with antenna array sparsity.

To determine how antenna length and optimal array pitch scale with wavelength, FDTD simulations of periodic antenna-coupled nanowire arrays were performed for varying antenna lengths and array pitches, with details provided in Section IV of the Supporting Information. In the simulations, a regime of moderately high ($\sim$ 40\% – 50\%) photon-absorption efficiency in the meander was observed over a range of array pitches at each antenna resonant wavelength, followed by a sharp drop in absorption efficiency as the array pitch was further increased. The optimal pitch is therefore defined as the largest pitch at which the nanowire photon-absorption efficiency at the antenna resonance exceeds 45\%, ensuring that the antenna array remains within the collective-coupling regime while maximizing antenna spacing. This 45\% threshold was chosen as a design criterion based on the FDTD results, representing a regime of near-maximum nanowire photon-absorption efficiency achievable across antenna resonant wavelengths from 4.5 µm to 7 µm for the device geometry considered.

The simulations also show that the antenna resonant wavelength ($\lambda_\mathrm{res}$) can be tuned by varying the antenna length ($L$), as shown in Figure \ref{fig:Fig4}c. In addition, the optimal array pitch increases with $\lambda_\mathrm{res}$ (Supporting Information, Figure S5e), so the number of antennas required in the periodic antenna array to span a 50 × 50 µ\si{m^2} geometric footprint decreases as $\lambda_\mathrm{res}$ increases. Since each antenna couples to a short nanowire segment (i.e., 80 nm-wide and 160 nm-long), the total photon-sensitive nanowire length decreases with increasing resonant wavelength. Thus, unlike a conventional meander, which maintains a fixed nanowire length for a given filling factor (e.g., 25\%), the periodic antenna-coupled nanowire meander at the optimal pitch can significantly reduce the total photon-sensitive nanowire length as the antenna resonant wavelength (i.e., detection wavelength) increases.

In the mid-IR regime, the noise-equivalent power (NEP) is often used as a performance metric for photodetectors, defined as $\text{NEP} = (h\nu/\eta)\sqrt{2D}$, where $h\nu$ is the photon energy, $\eta$ is the photon-detection efficiency and $D$ is the dark-count rate \cite{taylor2023low}.

We now compare the NEP of two SNSPD geometries in the mid-IR wavelength of 4.5 µm to 7 µm. The first is a conventional meandered SNSPD spanning 50 × 50 µ\si{m^2} area, consisting of an 80 nm-wide, 3 nm-thick WSi nanowire with 25\% filling fraction on a 300 nm \si{SiO_2} layer on Si substrate. FDTD simulations show that its photon-absorption efficiency under unpolarized illumination decreases from 2.1\% at 5 µm to 1.4\% at 7 µm (Supporting Information, Figure S2b). Assuming saturated internal detection efficiency, $\eta$ is therefore set by the photon-absorption efficiency. For the DCR estimation, we assume that the background-induced DCR scales almost linearly with the total nanowire length in the meander, as supported by Figure \ref{fig:Fig4}a. Hence, the background-induced DCR at the saturation knee current ($I_{\mathrm{sat}}$) is given by $D = N_{\mathrm{sq}} \times \Gamma_{\mathrm{sq}}$, where $\Gamma_{\mathrm{sq}}$ is the background-induced DCR per unit square length of the meander and $N_{\mathrm{sq}} = 9.7656 \times 10^4$ is the total number of squares of the photon-sensitive nanowire in the meander, determined from the total meander length set by the nanowire filling fraction. Here, $I_{\mathrm{sat}}$ denotes the bias current at which the SNSPD achieves saturated internal detection efficiency and is therefore often used as the operating bias. The resulting NEP of the meandered nanowire as a function of wavelength is shown in Figure \ref{fig:Fig4}d.

\begin{table}[t]
\caption{Comparison of the noise-equivalent power (NEP) of conventional and antenna-coupled nanowire meanders at $\lambda_{\mathrm{det}}$ = 5 µm and 7 µm. Both configurations span 50 × 50 µ\si{m^2} area with 80 nm-wide, 3 nm-thick WSi nanowires. Saturated internal detection efficiency and $\Gamma_{\mathrm{sq}} = 1 \times 10^{-3}$ Hz/$\square$ are assumed for both configurations.}
\label{tab:NEP_comparison}
\centering
\begin{tabular}{l c c c c c}
\hline
Configuration & $\lambda_{\mathrm{det}}$ & Nanowire filling fraction & $N_{\mathrm{sq}}$ & $\eta_{\mathrm{abs}}$ & NEP (W/$\sqrt{\text{Hz}}$)\\
\hline
Nanowire meander & 5 µm & 25\% & 97,656 & 2.08\% & $2.67\times10^{-17}$\\
\makecell[l]{Antenna-coupled\\nanowire meander} & 5 µm & \makecell[c]{$\sim 0.06\%$ \\ ($d$ = 4.51 µm)} & 246 & 45\% & $6.2\times10^{-20}$\\
\hline
Nanowire meander & 7 µm & 25\% & 97,656 & 1.42\% & $2.8\times10^{-17}$\\
\makecell[l]{Antenna-coupled\\nanowire meander} & 7 µm & \makecell[c]{$\sim 0.03\%$ \\ ($d$ = 6.63 µm)} & 121 & 45\% & $3.1\times10^{-20}$\\
\hline
\end{tabular}
\end{table}

The second configuration is the periodic antenna-coupled nanowire meander shown in Figure \ref{fig:Fig4}b, where each antenna is coupled to an 80 nm-wide, 160 nm-long WSi nanowire segment located on a \si{SiO_2} spacer above an Au reflector plane, with details provided in Section IV of the Supporting Information. The antenna length is chosen so that its resonant wavelength matches the target detection wavelength ($\lambda_{\mathrm{det}}$), and the optimal array pitch corresponding to the 45\% absorption criterion is used. The total number of squares of the photon-sensitive nanowire in the meander is determined by the number of antennas ($N_{\mathrm{ant}}$) required to span the 50 × 50 µ\si{m^2} area (Figure \ref{fig:Fig4}c). Since each antenna is coupled to two square-length nanowire segments, $N_{\mathrm{sq}}$ is given by $2 \times N_{\mathrm{ant}}$, and the background-induced DCR is given by $D = N_{\mathrm{sq}} \times \Gamma_{\mathrm{sq}}$. The resulting NEP values are shown in Figure \ref{fig:Fig4}d. The antenna-coupled meander achieves nearly three orders of magnitude lower NEP than the conventional meander because it provides higher photon-absorption efficiency with a much shorter total photon-sensitive nanowire length. \highlightblue{This improvement, however, represents an upper-bound estimate under well-shielded operating conditions with negligible background radiation within the antenna resonance bandwidth.}

Table \ref{tab:NEP_comparison} compares the NEP of the two configurations at $\lambda_{\mathrm{det}} = 5$ µm and 7 µm for $\Gamma_{\mathrm{sq}} = 1 \times 10^{-3}$ Hz/$\square$. In practice, $\Gamma_{\mathrm{sq}}$ depends mainly on the cryogenic device packaging and suppression of stray background photons. We estimate $\Gamma_{\mathrm{sq}} = 1.6 \times 10^{-2}$ Hz/$\square$ for our devices at a normalized bias current of 0.85, and $1.5 \times 10^{-3}$ Hz/$\square$ for the devices of Colangelo \textit{et al.} \cite{colangelo2022large} at the same normalized bias current, while values as low as $1 \times 10^{-6}$ Hz/$\square$ have been reported using a dark box \cite{taylor2023low}. Table \ref{tab:NEP_comparison} shows that antenna coupling reduces the NEP by approximately 430× at 5 µm and 900× at 7 µm relative to a conventional meander of the same detection area, \highlightblue{under well-shielded operating conditions with negligible background radiation near the antenna resonance}. The resulting NEP, on the order of $3 \times 10^{-20}$ W/$\sqrt{\text{Hz}}$, meets the sensitivity requirements of flagship-class infrared space telescopes \cite{taylor2023low} and is promising for dark-matter search experiments \cite{koppell2026dark}. Overall, these results indicate that antenna-coupled nanowire meanders provide a scalable route to increasing the photon-detection area of SNSPDs while maintaining low-NEP performance in the mid-IR and far-IR regimes.

\vspace{-0.3cm}
\section{Conclusions}
\vspace{-0.3cm}
In conclusion, we demonstrated photon-absorption enhancement in a 3 nm-thick, 80 nm-wide WSi SNSPD at 7.4 µm via resonant metallic antenna coupling. By concentrating incident radiation at the feed gap, a crossed bowtie antenna increased the effective photon-absorption area of a short WSi nanowire segment by 15.7× relative to an otherwise identical bare nanowire, while maintaining similar dark-count rates. The antenna-induced field enhancement is strongly localized near the feed gap, indicating that short photon-sensitive nanowires are preferred to maximize photon-absorption enhancement while minimizing dark counts from nanowire segments outside the antenna-enhanced region. This localized coupling motivates periodic antenna-coupled nanowire arrays as a route to large-area scaling. Our analysis indicates that, \highlightblue{under well-shielded operating conditions with negligible resonant background radiation,} such arrays can reduce the NEP by approximately 900× at 7 µm for 50 × 50 µ\si{m^2} footprint relative to a conventional meandered nanowire of the same geometric footprint. Our results highlight antenna-coupled nanowire arrays as a promising route to realizing large-area, low-NEP, high-efficiency SNSPDs for single-photon detection in infrared astronomy, dark-matter detection, and biological sensing at mid- and far-infrared wavelengths.

\vspace{-0.3cm}
\section*{Supporting Information}
\vspace{-0.3cm}
Additional details on FTIR-based refractive index characterization, FDTD simulation framework, dependence of photon-absorption enhancement on antenna-coupled nanowire length and antenna feed-gap alignment, resonant-wavelength tunability and array-pitch optimization in periodic antenna-coupled nanowires, device fabrication, and mid-infrared SNSPD measurements are provided.

\vspace{-0.3cm}
\section*{Acknowledgements}
\vspace{-0.3cm}
This work was carried out in part using the facilities at MIT.nano, and the authors gratefully acknowledge Mark Mondol, James Daley, and Juan Ferrera for their technical support in device nanofabrication. D.J.P. also thanks Marco Colangelo for helpful discussions during the early stages of this work. The Lumerical FDTD simulations were performed on the MTL CAD server at MIT. Part of this research was also carried out at the Jet Propulsion Laboratory, California Institute of Technology, under a contract with the National Aeronautics and Space Administration (80NM0018D0004). The research was supported by NASA under the ROSES-APRA program. D.J.P. acknowledges support from the MathWorks Engineering Fellowship. The authors also thank Malick Sere, Daniel Graham, and Yashika Kapoor for their feedback during the preparation of the manuscript.

\vspace{-0.3cm}
\section*{Contributions}
\vspace{-0.3cm}
D.J.P., S.K., K.K.B., P.D.K., and B.K. conceived the idea. D.J.P. carried out the FDTD simulations, performed numerical analysis, designed the device layout, and executed the device fabrication. S.K. contributed to the device design and numerical analysis. A.D.B. and S.P. were responsible for the WSi film deposition, while G.G.T., B.K., and S.P. conducted the cryogenic measurements. D.J.P. wrote the manuscript with input from all authors. B.K., E.E.W., M.D.S., P.D.K., and K.K.B. supervised the project.

\bibliography{mid_antenna_ref}

\newpage
\section*{TOC Graphic}
\vspace{0.3cm}
\begin{figure}[!htbp]
    \vspace{-0.5cm}
    \centering
    \includegraphics[width=\textwidth]{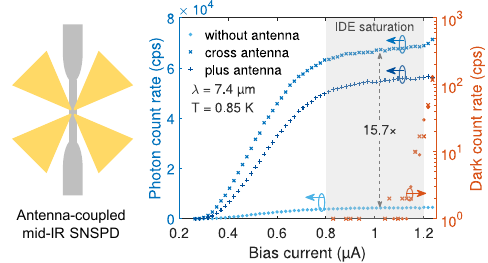}
    \vspace{-0.5cm}
    \label{fig:TOC_Fig}
\end{figure}

\end{document}


\maketitle
\newpage
\tableofcontents
 
\renewcommand{\thefigure}{S\arabic{figure}}
\setcounter{figure}{0} 

\vspace{-0.3cm}
\subsection*{I. FTIR-Based Refractive Index Characterization}
\begin{figure}[!t]
    \centering
    \includegraphics[width=0.95\textwidth]{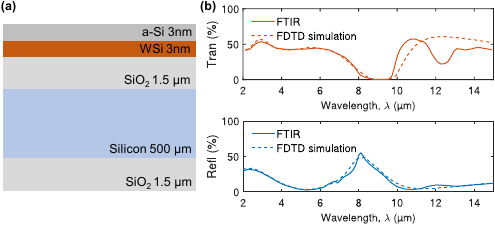}
    \vspace{-0.2cm}
    \caption{\small \textbf{(a)} A schematic of the material stack in the WSi thin-film sample used in this work is shown. \textbf{(b)} The FTIR transmittance of the sample was measured at a near-normal incidence angle (\(\sim\)12°), while the reflectance was obtained using an FTIR microscope with a numerical aperture of 0.58. These measurements were then compared with FDTD simulations for the same material stack, confirming the accuracy of the refractive-index models used in the simulations.\vspace{-3mm}}
    \label{fig:FigS1}
\end{figure}

\addcontentsline{toc}{subsection}{I. FTIR-Based Refractive Index Characterization}
\vspace{-0.3cm}
In this work, we performed finite-difference time-domain (FDTD) simulations in Ansys Lumerical software to optimize the geometric dimensions of the antenna structures. In the FDTD simulations, we used Palik’s refractive index database \cite{palik1998handbook} for the optical properties of silicon, \si{SiO_2}, and gold, while the refractive index of WSi was obtained from the measurements performed by our collaborators at JPL. To evaluate whether these refractive indices accurately represent the optical properties of the WSi sample, we performed FTIR reflection and transmission measurements on a WSi-coated sample, as shown in Figure \ref{fig:FigS1}a, using a Thermo Scientific Nicolet iS50 Fourier Transform Infrared Spectrometer (FTIR).

The FTIR transmission measurement was performed at a near-normal incidence angle, and the source spectrum was used to normalize the sample’s transmitted spectrum. The reflection measurement was performed using an FTIR microscope in a reflection configuration with a numerical aperture of 0.58, and the reflected spectrum of a gold mirror was used to normalize the sample’s reflection spectrum. Both transmission and reflection measurements were performed with an unpolarized collimated beam, and each measurement was averaged over 64 scans with automatic atmospheric suppression enabled.

We then simulated the transmission and reflection spectra for the same material stack, taking into account the source polarization and incidence angle, and compared the results with the FTIR measurements, as shown in Figure \ref{fig:FigS1}b. The simulated data show good agreement with the FTIR results, thereby validating the refractive index data used in our simulations. The refractive indices of these materials could be further refined using FTIR reflection measurements at various incidence angles \cite{paul2025determination}, although we did not perform such measurements in this work.

\vspace{-0.3cm}
\subsection{II. FDTD Simulation Framework}
\begin{figure}[!t]
    \vspace{-0.5cm}
    \centering
    \includegraphics[width=0.75\textwidth]{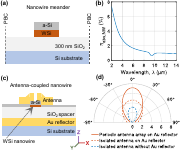}
    \vspace{-0.2cm}
    \caption{\small \textbf{(a)} Schematic of the FDTD simulation setup for WSi nanowire meander on 300 nm \si{SiO_2}/Si substrate with periodic boundary conditions (PBCs) is shown. \textbf{(b)} FDTD simulation results showing the photon-absorption efficiency ($\eta_{abs,NW}$) of the WSi nanowire meander plotted as a function of photon wavelength. \textbf{(c)} Schematic cross-section of an antenna-coupled WSi nanowire is shown. For the periodic antenna array configuration, PBCs are applied at the chosen array pitch. \textbf{(d)} Antenna directivity at the resonant wavelength for different configurations is plotted. Directivity increases with the inclusion of the Au reflector plane and with the periodic antenna array configuration, enhancing the effective photon-absorption area of the nanowire. \vspace{-3mm}}
    \label{fig:FigS2}
\end{figure}

\addcontentsline{toc}{subsection}{II. FDTD Simulation Framework}
\vspace{-0.3cm}
In this work, we performed finite-difference time-domain (FDTD) simulations to investigate both bare WSi nanowire meanders and antenna-coupled nanowire geometries. For the bare nanowire meander configuration, as shown in Figure \ref{fig:FigS2}a, we performed two-dimensional FDTD simulations of a 3 nm-thick WSi nanowire on a 300 nm-thick \si{SiO_2} layer on Si substrate to calculate the photon-absorption efficiency of the WSi nanowire as a function of wavelength. Periodic boundary conditions (PBCs) were applied in the lateral directions to represent the meandered configuration with a nanowire filling factor of 25\%, while perfectly matched layers (PMLs) were used along the light-propagation direction. For an unpolarized source at normal incidence, the simulated photon-absorption efficiency monotonically decreases from 2\% at 5 µm to 1\% at 8 µm, as shown in Figure \ref{fig:FigS2}b. 

For the antenna-coupled nanowire geometry, three-dimensional FDTD simulations were performed. Figure \ref{fig:FigS2}c shows the schematic cross-section of an antenna-coupled nanowire, including the antenna, WSi nanowire, a-Si encapsulation layer, \si{SiO_2} spacer, and Au reflector plane on a Si substrate. In the FDTD simulations, the Au reflector was modeled as a perfect electric conductor (PEC), which is a good approximation for a thick Au back-reflector in the mid-IR. For simulations of periodic antenna arrays, PBCs were applied at the specified array pitch, whereas isolated antenna simulations employed PML boundary conditions in all directions. To minimize the computational cost of the three-dimensional FDTD simulations while retaining mesh accuracy in regions of strong field confinement, a nonuniform mesh was used in the antenna-coupled nanowire simulations, particularly within the antenna feed gap and near the nanowire. The mesh grid size was systematically increased until the near-field amplitude spectra converged, with further increases producing negligible changes in the results. In the FDTD simulations, a coherent plane-wave source was used, although the experiment uses a spectrally filtered incoherent thermal source. As the antenna behaves as a linear passive element, its field-enhancement depends on the local structural resonance rather than on the temporal coherence of the source. Given that the experimental thermal source is filtered to a narrow bandwidth centered at the antenna resonance, the coherent-source FDTD simulations should yield a highly reliable estimate of the experimental field enhancement for the incoherent thermal source.

To investigate the angular radiation characteristics and effective photon-collection capability of the antenna, we simulated the far-field directivity of the antenna using the directivity monitor in Lumerical. The antenna's effective aperture depends on its directivity and is given by $A_{\mathrm{eff}} \approx \lambda^2 D_{\mathrm{ant}} / (4\pi)$ \cite{eggleston2015optical,balanis2015antenna}, where $\lambda$ is the resonant wavelength and $D_{\mathrm{ant}}$ is the antenna directivity. The antenna array at the optimal periodicity exhibits higher directivity compared to an isolated antenna, which increases the antenna aperture ($A_{\mathrm{eff}}$), or equivalently, the effective photon-absorption area of the nanowire in the array configuration. Figure \ref{fig:FigS2}d illustrates how the peak directivity varies with the antenna geometry. For a single antenna without a metallic ground plane, the antenna directivity ($D_{\mathrm{ant}}$) is 3.13. When a metallic ground plane is added, the directivity increases to 4.6. The directivity further increases to 7.15 in the periodic antenna array configuration. These results indicate that a periodically arranged antenna array on a metallic ground plane provides the most effective configuration for enhancing photon absorption in SNSPDs.
\begin{figure}[!b]
    \centering
    \includegraphics[width=0.95\textwidth]{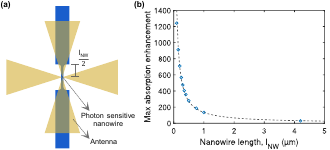}
    \vspace{-0.2cm}
    \caption{\small \textbf{(a)} Schematic of an antenna-coupled nanowire geometry, with the photon-sensitive nanowire length centered on the antenna feed gap. \textbf{(b)} Maximum photon absorption enhancement factor as a function of the photon-sensitive nanowire length ($l_{\mathrm{NW}}$), obtained from FDTD simulations by calculating the ratio of optical power absorbed in the antenna-coupled nanowire to that in an identical nanowire without an antenna. The circled points represent FDTD simulation results, while the dashed line shows a power-law fit to the data. The enhancement factor decreases with increasing nanowire length and follows a power-law decay, indicating that the antenna-induced field enhancement is strongly localized near the feed gap, with only a short portion of the nanowire effectively interacting with the antenna near-field. \vspace{-3mm}}
    \label{fig:FigS3}
\end{figure}

\vspace{-0.3cm}
\subsection{III. Dependence of the Observed Photon-Absorption Enhancement on Nanowire Length and Antenna Feed-Gap Alignment}
\addcontentsline{toc}{subsection}{III. Dependence of the Observed Photon-Absorption Enhancement on Nanowire Length and Antenna Feed-Gap Alignment}
\vspace{-0.3cm}
The observed photon absorption enhancement in an antenna-coupled SNSPD depends on both the antenna geometry and the length of the photon-sensitive nanowire coupled to the antenna. As discussed in the main text, the electromagnetic field enhancement produced by the antenna is highly localized near the antenna feed gap, so only a short segment of the nanowire experiences strong field enhancement and contributes to increased photon absorption efficiency. When the nanowire length extends beyond this localized region, the remaining nanowire length absorbs photons similarly to an identical nanowire without antenna coupling. As a result, the overall enhancement factor, defined relative to an identical nanowire without the antenna, decreases with increasing nanowire length. This length dependence explains why the simulated enhancement for the 200 nm-long nanowire in Figure 1d is much larger than that obtained for the fabricated 4.2 µm-long nanowire in Figure 2c, and underscores why short photon-sensitive nanowires are preferred in antenna-coupled SNSPDs when scaling to larger detector footprint.

We performed FDTD simulations for the antenna-coupled nanowire geometry shown in Figure \ref{fig:FigS3}a and compared it to an identical nanowire without an antenna. The simulated structure uses the same material stack as the device shown in Figure 2a of the main text, consisting of a cross-oriented Au bowtie antenna coupled to a 3 nm-thick, 80 nm-wide WSi nanowire, with 7 nm-thick a-Si encapsulation layer between the nanowire and antenna and 1.5 µm-thick \si{SiO_2} spacer on the silicon substrate. In this geometry, the center of the photon-sensitive nanowire is aligned with the center of the antenna feed gap, and the total nanowire length ($l_\mathrm{NW}$) is varied while all other antenna dimensions and material parameters are kept fixed. Figure \ref{fig:FigS3}b shows the simulated absorption-enhancement factor at the antenna resonant wavelength as a function of $l_\mathrm{NW}$, where the enhancement factor is defined as the ratio of the photon absorption in the antenna-coupled nanowire to that in an otherwise identical nanowire without the antenna. The enhancement factor decreases monotonically with increasing nanowire length and follows a power-law dependence. This behavior reflects the highly localized antenna near-field at the feed gap, such that only a short region of the nanowire experiences strong antenna-induced field concentration, while the remaining nanowire absorbs similarly to a bare nanowire. As a result, increasing the nanowire length gradually reduces the average enhancement factor. This trend explains the larger enhancement predicted for the 200 nm-long nanowire in Figure 1d compared with the smaller enhancement observed for the 4.2 µm-long device in Figure 2c and supports the use of short photon-sensitive nanowires in antenna-coupled SNSPDs.

\begin{figure}[!b]
    \centering
    \includegraphics[width=0.97\textwidth]{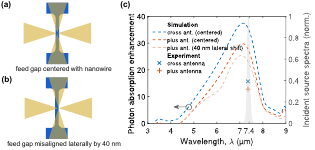}
    \vspace{-0.2cm}
    \caption{\small \textbf{(a)} Schematic of a plus-oriented bowtie antenna with the 40 nm-wide antenna feed gap centered on an 80 nm-wide nanowire. \textbf{(b)} Schematic of the same structure with the antenna feed gap laterally shifted by 40 nm, placing the feed-gap center at the nanowire edge. \textbf{(c)} Simulated photon-absorption enhancement for the centered and 40 nm laterally shifted configurations, together with the simulated spectrum of the centered cross-oriented antenna. Experimentally measured enhancement factors for the cross- and plus-oriented antennas are shown for comparison. The shaded gray region indicates the normalized spectrum of the 7.4 µm illumination source used in the measurements. \vspace{-6mm}}
    \label{fig:FigS3b}
\end{figure}
The photon-absorption enhancement also depends on the alignment between the antenna feed gap and the photon-sensitive nanowire. In this work, the antenna feed gap is 40 nm wide and the nanowire is 80 nm wide, so centering the feed gap on the nanowire requires high overlay precision. In the fabricated devices, the overlay accuracy between the antenna feed-gap center and the nanowire center was within ±10 nm. To evaluate the effect of fabrication-induced misalignment on the observed enhancement factor, we performed FDTD simulations for a plus-oriented antenna with the feed gap centered on the nanowire and compared the results with a configuration in which the feed gap was laterally shifted by 40 nm, placing the feed-gap center at the nanowire edge (Figures \ref{fig:FigS3b}a, \ref{fig:FigS3b}b). The simulated enhancement spectra are shown in Figure \ref{fig:FigS3b}c. A 40 nm lateral misalignment reduces the enhancement factor by approximately 15\% at the resonant wavelength, indicating that the enhancement is sensitive to the overlap between the localized antenna near field and the nanowire. Nevertheless, even with this large misalignment, the simulated enhancement of the plus-oriented antenna remains higher than the experimentally measured value, indicating that additional factors contributed to the observed discrepancy. This alignment sensitivity is expected to become less critical at longer resonant wavelengths because the antenna dimensions, including the feed-gap width, increase with wavelength, thereby relaxing fabrication tolerances.

\vspace{-0.3cm}
\subsection{IV. Resonant Wavelength Tunability and Array Pitch Optimization in Periodic Antenna-Coupled Nanowires}
\addcontentsline{toc}{subsection}{IV. Resonant Wavelength Tunability and Array Pitch Optimization in Periodic Antenna-Coupled Nanowires}
\vspace{-0.3cm}
While coupling a resonant antenna to a photon-sensitive nanowire increases its effective photon-detection area beyond its geometric footprint, the resulting detection area of a single antenna-coupled nanowire may still be insufficient for applications requiring large detection areas. Hence, to scale the antenna-coupled nanowire configuration to large detection-area footprint, we examined periodic arrangement of antenna-coupled nanowires using FDTD simulations. In this periodic array configuration, short photon-sensitive antenna-coupled nanowires are connected in series by relatively wide photon-insensitive superconducting strips, forming a meander-like geometry.

\begin{figure}[!htbp]
    \vspace{-0.9cm}
    \centering
    \includegraphics[width=0.97\textwidth]{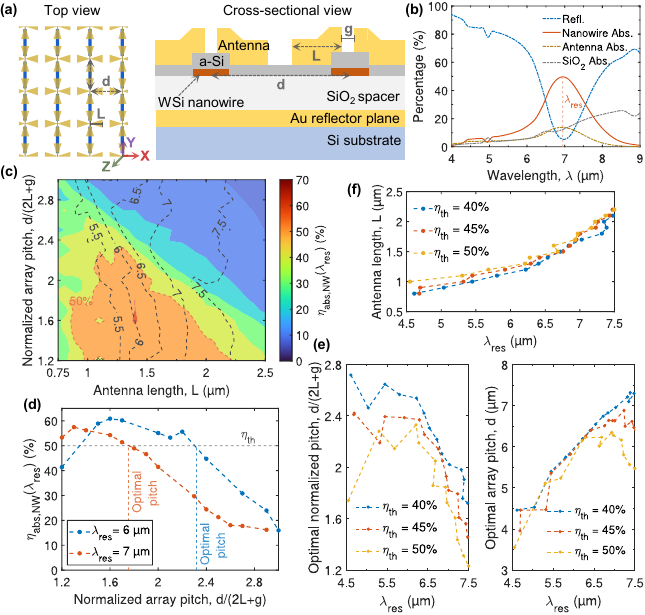}
    \vspace{-0.4cm}
    \caption{\small Scaling SNSPD photon-detection area with periodic antenna arrays. \textbf{(a)} A periodic antenna array configuration, featuring plus-oriented bowtie antennas on 80 nm-wide, 160 nm-long superconducting nanowire segments connected by tapered 500 nm-wide wires in a meandered geometry, is shown. The antenna length is denoted by $L$ and the array pitch by $d$. The array pitch is kept identical in the x- and y-directions. The cross-sectional view shows the Au reflector plane and \si{SiO_2} as the spacer oxide. \textbf{(b)} FDTD simulation results of a periodic array of antenna-coupled nanowires with $L$ = 1.8 µm and $d$ = 6.2 µm show that at the antenna resonant wavelength ($\lambda_\mathrm{res}$ = 6.95 µm), 50\% of incident photons are absorbed by the 80 nm-wide, 160 nm-long superconducting nanowire segments, while 14\% are absorbed by the antenna and 5\% are reflected. \textbf{(c)} The photon-absorption efficiency of the nanowire at the resonant wavelength, $\eta_{abs,NW}(\lambda_\mathrm{res})$, is shown as a function of antenna length ($L$) and normalized array pitch, with contour lines indicating $\lambda_\mathrm{res}$ in µm. The antenna feed gap ($g$) is kept fixed at 40 nm, and the spacer-oxide thickness is 1 µm. \textbf{(d)} Nanowire photon-absorption efficiency at the resonant wavelength, $\eta_{abs,NW}(\lambda_\mathrm{res})$, is plotted as a function of the normalized array pitch for $\lambda_\mathrm{res}$ = 6 µm and 7 µm. Vertical dashed lines indicate the corresponding optimal array pitch for an absorption-efficiency threshold ($\eta_{\mathrm{th}}$) of 50\%. \textbf{(e)} Optimal normalized array pitch and absolute array pitch are plotted as a function of the resonant wavelength ($\lambda_\mathrm{res}$) for different nanowire absorption-efficiency thresholds, $\eta_{\mathrm{th}}$. \textbf{(f)} Optimal antenna length ($L$) is plotted as a function of the resonant wavelength for different nanowire absorption-efficiency thresholds, showing the tunability of $\lambda_\mathrm{res}$ with antenna length.}
    \label{fig:Fig3}
\end{figure}

Figure \ref{fig:Fig3}a shows a schematic of a periodic antenna-coupled nanowire array with antenna array pitch $d$. In this schematic, we consider plus-oriented bowtie antennas; however, cross-oriented antennas can also be used. As shown in the cross-sectional view in Figure \ref{fig:Fig3}a, the array pitch sets the spacing between neighboring photon-sensitive nanowire segments. Since each antenna is coupled to a photon-sensitive nanowire segment, appropriately spaced antennas can capture more photons while requiring a shorter total photon-sensitive nanowire length, making this configuration less susceptible to background-induced dark-count rates than a meandered SNSPD of similar geometric footprint while achieving higher photon-count rates. Therefore, the array pitch should be optimized to maximize the collective resonance and photon-absorption efficiency while minimizing the total length of photon-sensitive nanowires within a given footprint.

Figure \ref{fig:Fig3}b shows FDTD simulation results for a periodic array of antenna-coupled nanowires with an array pitch of 6.2 µm. At the resonant wavelength, 50\% of incident 6.95 µm photons are absorbed by the 80 nm-wide, 160 nm-long nanowire segments, while 5\% is reflected. The remaining incident photons are absorbed by the antennas, the a-Si encapsulation layer, and the \si{SiO_2} spacer oxide. Overall, the simulation shows that antenna coupling enables a short nanowire segment (80 nm $\times$ 160 nm) to absorb 50\% of the photons incident on a 6.2 µm $\times$ 6.2 µm area. This reduction in the total photon-sensitive nanowire length, compared to a conventional meandered nanowire spanning 6.2 µm $\times$ 6.2 µm area, while maintaining high photon-absorption efficiency, makes the antenna-coupled nanowire configuration well-suited for scaling to larger detection areas.

To examine how the array pitch ($d$) and antenna arm length ($L$) affect the collective antenna resonance in a periodic array configuration, we performed parametric FDTD simulations by varying these two parameters. In the simulations, an Au reflector plane beneath the dielectric spacer was considered, with the \si{SiO_2} spacer thickness fixed at 1 µm. We considered plus-oriented quadrupole bowtie antennas coupled to 80 nm-wide, 160 nm-long WSi nanowires connected by 500 nm-wide superconducting stripes. The WSi thickness was 3 nm, with 7 nm a-Si encapsulation layer to isolate it from the antenna. The antenna feed gap ($g$) was fixed at 40 nm, the Au antenna thickness at 45 nm, and the antenna arm width ($h$) was set to 0.642$L$. Each nanowire segment was kept at 160 nm, since we observed in the FDTD simulations that the antenna-enhanced near-field is primarily confined to the vicinity of the antenna feed gap. Hence, extending the nanowire length beyond a few feed-gap lengths (i.e., on the order of 2--5$\times$) does not significantly increase the photon-absorption efficiency in the nanowire, but instead increases its physical footprint and thereby the DCR.

Figure \ref{fig:Fig3}c shows how the resonant wavelength ($\lambda_\mathrm{res}$) and the photon-absorption efficiency of the nanowire at the resonant wavelength ($\eta_{abs,NW}(\lambda_\mathrm{res})$) vary with the antenna arm length ($L$) and the normalized array pitch. Since the antenna arm length was varied for each array pitch, it was convenient to plot the normalized array pitch as the y-axis in this figure, with a normalized value of 1 corresponding to an array pitch, $d = 2L + g$. As indicated by the contour lines in Figure \ref{fig:Fig3}c, the resonant wavelength depends strongly on the antenna arm length ($L$) and relatively weakly on the array pitch. Therefore, varying the antenna arm length in a periodic antenna array enables tuning of the resonant wavelength. This tunability can be exploited to design multispectral mid-IR SNSPD arrays on a single substrate, in which different detector pixels on the same chip are optimized for different wavelengths. In this configuration, each antenna array is designed to have a distinct resonance wavelength and is coupled to photon-sensitive superconducting nanowires to detect photons at that resonance wavelength with high photon-absorption efficiency. Overall, this wavelength tunability provides a versatile platform for designing complex mid-IR spectroscopic and imaging systems using antenna-coupled SNSPDs.

In addition, Figure \ref{fig:Fig3}c shows the FDTD-simulated photon-absorption efficiency of periodically spaced antenna-coupled nanowires as a function of antenna length and pitch. The peak value of $\eta_{abs,NW}(\lambda_\mathrm{res})$ is approximately 60\% at an antenna length of 1.4 µm and normalized pitch of 1.6 -- 1.8, although this high-efficiency region is narrow and observed only near an antenna resonant wavelength of 6 µm, primarily because \si{SiO_2} spacer thickness is fixed at 1 µm in this simulation sweep. A broader regime of moderately high absorption efficiency ($\sim$ 40\% -- 60\%) is observed over a range of array pitches at multiple resonant wavelengths (i.e., 4.5 -- 7.5 µm), followed by a sharp decrease as the array pitch increases further. 

In the periodic array configuration, long-range dipolar interactions among the antennas dominate the collective resonance behavior, causing the nanowire absorption efficiency to remain nearly constant over a certain pitch range. As the array pitch increases further, long-range dipolar interactions among the antennas weaken, causing the antenna array behavior to approach that of an isolated antenna and the absorption efficiency to drop. Therefore, the array pitch should be chosen to balance high photon absorption in the nanowire with increased antenna spacing, thereby reducing the total length of photon-sensitive nanowire segments required to cover a given area. We define this pitch as the optimal array pitch, corresponding to the largest pitch at which the array remains in the collective-coupling regime while maximizing antenna spacing.

Figure \ref{fig:Fig3}d shows the photon-absorption efficiency $\eta_{abs,NW}(\lambda_\mathrm{res})$ at two resonant wavelengths as a function of normalized array pitch. A plateau of moderately high absorption ($>$ 40\%) is observed over a range of shorter normalized array pitches, followed by a rapid drop as the pitch increases, indicating a transition from a collective-coupling regime to an isolated-antenna regime. To quantitatively define the optimal array pitch, we introduce a design criterion called the absorption-efficiency threshold ($\eta_{\mathrm{th}}$), which balances maximizing the antenna spacing while remaining within the collective-coupling plateau. The optimal array pitch for $\lambda_\mathrm{res}$ = 6 µm and 7 µm at $\eta_{\mathrm{th}} = 50\%$ is shown in Figure \ref{fig:Fig3}d. The choice of $\eta_{\mathrm{th}}$ is a design parameter, and Figure \ref{fig:Fig3}e shows the optimal array pitch as a function of $\lambda_\mathrm{res}$ for $\eta_{\mathrm{th}}$ values of 40\%, 45\%, and 50\%. Lower $\eta_{\mathrm{th}}$ values yield larger optimal pitches, highlighting a trade-off between antenna density (i.e., periodicity) and nanowire absorption efficiency in the periodic antenna-coupled nanowire meander configuration. In addition, we observe in Figure \ref{fig:Fig3}c that $\eta_{abs,NW}(\lambda_\mathrm{res})$ does not reach 50\% or higher at longer resonant wavelengths ($\lambda_\mathrm{res} >$ 7.5 µm) for the simulated device geometry, due to the \si{SiO_2} spacer thickness being fixed at 1 µm. The optimal array pitch will vary with the spacer oxide thickness; however, it is not examined here, as including spacer oxide thickness as a third parameter in the FDTD simulation sweep, along with antenna length and array pitch, would make the simulations time-consuming. Instead, since the motivation for this array-pitch optimization analysis is to understand how the number of antennas in a periodic antenna-array configuration scales with antenna resonant wavelength in the mid-infrared and to evaluate the resulting SNSPD performance of an antenna-coupled meander relative to a conventional nanowire meander, we kept the spacer oxide thickness fixed and varied only the antenna length and array pitch.

Finally, Figure \ref{fig:Fig3}f shows the optimal antenna length ($L$) as a function of the resonant wavelength for different nanowire absorption-efficiency thresholds, highlighting the tunability of $\lambda_\mathrm{res}$ with antenna length. The slight nonlinearity in the antenna-length scaling with resonant wavelength arises from the antenna array operating in the collective-resonance regime. In addition, the spacer oxide thickness, fixed at 1 µm in this simulation, also influenced the antenna resonance condition.

\vspace{-0.3cm}
\subsection{V. Device Fabrication}
\addcontentsline{toc}{subsection}{V. Device Fabrication}
\vspace{-0.3cm}
The device fabrication was carried out in three lithography steps at the MIT.nano cleanroom. In the first step, alignment markers and contact pads were fabricated on a 1$\times$1 \si{cm^2} WSi sample using photolithography, followed by a Ti-Au liftoff process.

In the second lithography step, the WSi nanowires were patterned by electron-beam lithography (EBL), followed by reactive ion etching. The sample was spin-coated with maN-2401 negative-tone resist at 3k rpm and baked at 90°C for 1 minute. EBL was performed using an Elionix HS-50 system with a base dose of 465 \si{\micro\coulomb}/\si{cm^2} with proximity-effect correction. The exposed sample was developed in AZ726 for 15 s, rinsed in DI water for 60 s, and dried with nitrogen. WSi etching was carried out with \si{CF_4} plasma at 30 W bias power. The resist was lifted off by immersing the sample in NMP at 60°C for a few hours. Following this process, we were able to achieve 80 nm-wide WSi nanowires.

During early process development, we observed persistent maN-2401 residue on the patterned nanowires after etching, which NMP alone could not fully remove. This was mitigated by adopting a segmented slow-etch process to prevent resist overheating. A short oxygen plasma clean could also remove the residue; however, we did not use ashing in this work. The residue issue was more severe with maN-2401 than with ZEP-520 positive-tone resist under identical etch conditions. Nonetheless, the negative-tone resist was essential in this step to remove the WSi beneath the antenna region and avoid unwanted absorption loss.
\begin{figure}[!t]
    \vspace{-0.5cm}
    \centering
    \includegraphics[width=0.95\textwidth]{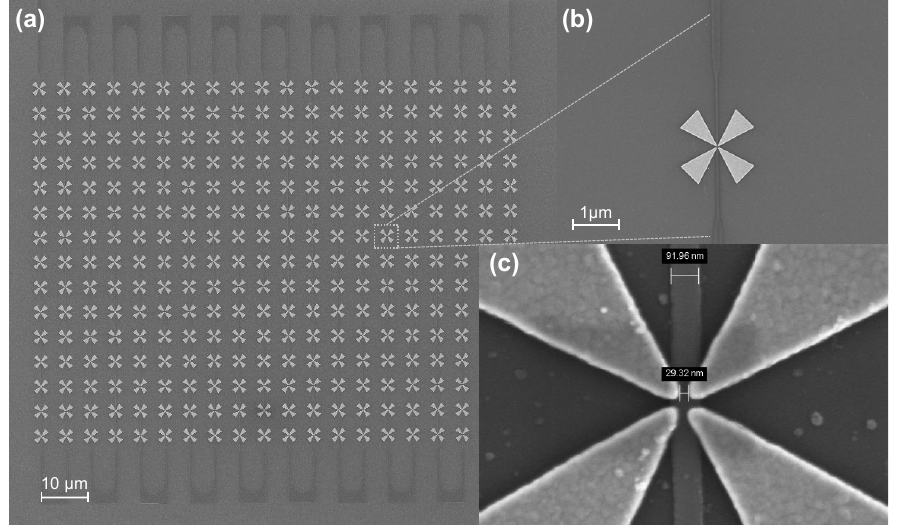}
    \vspace{-0.1cm}
    \caption{\small \textbf{(a)} A scanning electron microscopy (SEM) image of a fabricated crossed-bowtie antenna array on a 100 $\times$ 100 \si{\micro\meter\squared} nanowire meander is shown, demonstrating successful alignment of the antenna feedgaps with the nanowire centers over a large active area. The array pitch is \SI{5.5}{\micro\meter}. \textbf{(b)} A single antenna from the array is highlighted. \textbf{(c)} A zoomed-in view of that antenna, showing a 30 nm antenna feedgap aligned with the 90-nm-wide nanowire.\vspace{-3mm}}
    \label{fig:FigS4}
\end{figure}

In the final lithography step, bowtie antennas were fabricated using EBL, followed by a Ti-Au liftoff process. Prior to this, approximately 4 nm of a-Si was deposited at room temperature using an AJA magnetron sputtering tool to ensure electrical and thermal isolation between the WSi nanowire and the Au antenna. Without this layer, the Au antenna could act as an electrical and thermal shunt for the nanowire when it transitions to the resistive hotspot state after photon absorption, potentially affecting the SNSPD’s reset dynamics.

After a-Si deposition, the sample was coated with anisole-diluted ZEP-530A (1:1) positive-tone resist at 3k rpm and baked at 180°C for 2 min. The measured resist thickness was around 90 nm, which performed better for achieving sub-50 nm antenna feed gaps compared to undiluted ZEP-530A. Electra 92 (AR-PC 5090) was then spin-coated at 5k rpm and baked at 90°C for 1 minute as a charge-dissipation layer. This charge dissipation layer helped to achieve the desired antenna shape and feed-gap between adjacent bowtie tips, thereby improving the overall antenna liftoff yield. EBL was carried out in the HS-50 system with a base dose of 275 \si{\micro\coulomb}/\si{cm^2} and proximity-effect correction. Auto-alignment during the EBL step enabled consistent alignment accuracy within 5 nm between the antenna and the underlying nanowires. This high overlay precision was achieved using the automatic alignment-marker search routine of the Elionix HS-50 EBL system together with a hierarchical marker layout. Four cross-shaped Au markers near the chip corners were used for global alignment, while local cross-shaped Au markers within each antenna write field were used immediately before antenna exposure to correct residual translational and rotational errors. The automatic alignment procedure was iterated until the residual marker-position error was below 5 nm. Clean, continuous Au markers with well-defined lift-off edges were found to be essential for reliable marker detection and reproducible antenna-to-nanowire overlay alignment across the large-area meander, as shown in Figure \ref{fig:FigS4}.

Following exposure, Electra 92 was removed by DI water rinse, and the sample was developed in o-xylene at \SI{-5}{\celsius} for 30 s, rinsed in IPA for 30 s, and dried with nitrogen. Next, 2 nm Ti followed by 45 nm Au was deposited by e-beam evaporation, and the resist was lifted off in NMP at 60 °C for a few hours. In this fabrication run, both antenna-coupled short-nanowire devices and a large-area antenna-coupled nanowire meander were fabricated, and representative SEM images are shown in Figure \ref{fig:FigS4}. The antenna array was patterned on a 100 $\times$ 100 \si{\micro\meter\squared} nanowire meander, demonstrating the lithographic capability of aligning the antenna feedgaps with the nanowire centers over a large area. However, this antenna-coupled 100 $\times$ 100 \si{\micro\meter\squared} meandered SNSPD was electrically open at room temperature, likely due to a combination of fabrication-induced defects and local material imperfections in the 3-nm-thick WSi film. Therefore, the experimental data presented in the main text are limited to the successfully fabricated antenna-coupled short-nanowire devices.

\vspace{-0.3cm}
\subsection{VI. Mid-IR SNSPD Measurements}
\begin{figure}[!b]
     \centering
     \includegraphics[width=0.85\textwidth]{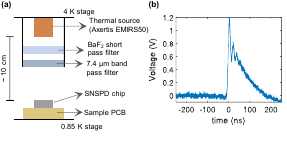}
     \vspace{-1.1cm}
     \caption{\small \textbf{(a)} The mid-IR source setup in the cryostat is shown. The thermal source and filters were enclosed in a 0.5 inch diameter lens tube, with shortpass and bandpass optical filters placed in front of the source, and the lens tube attached to the 4 K stage. \textbf{(b)} The output voltage pulse from an antenna-coupled nanowire for \SI{7.4}{\micro\meter}-wavelength photon is shown. The threshold voltage in the counter was set to 1 V to count pulses in the photon-count rate (PCR) and dark-count rate (DCR) measurements. The PCR and DCR are shown in Figure 2b. \vspace{-3mm}}
     \label{fig:FigS5}
\end{figure}

\addcontentsline{toc}{subsection}{VI. Mid-IR SNSPD Measurements}
\vspace{-0.3cm}
To illuminate the SNSPDs with mid-IR photons, a broadband thermal blackbody source (Axertis EMIRS50) was mounted on the 4 K stage of the cryostat, as shown in Figure \ref{fig:FigS5}a. The thermal source includes an aluminum reflector that concentrates about 80\% of the emission power into 20° angle at a working distance of 15 -- 30 mm, with peak emissivity ($\varepsilon$) of 0.85. The WSi SNSPD chip was mounted on the 0.85 K base-temperature plate of the cryostat with approximately 10 cm separation between the thermal source and the chip. The spectral emission power of the source was electrically modulated to keep the emitted power low enough to maintain the cryostat base temperature at 0.85 K during operation.

To configure the mounted thermal source as a \SI{7.4}{\micro\meter} mid-IR emitter, we placed a shortpass filter followed by a narrow bandpass filter in front of the thermal source. We used a 3 mm barium fluoride window (Thorlabs WG00530-E) as the shortpass filter, which exhibits strong absorption above \SI{12}{\micro\meter}, with an AR coating that further reduces transmission above \SI{8}{\micro\meter}. A Spectrogon NB-7400-148-nm filter was used as the bandpass filter, with a central wavelength of 7400 nm $\pm$ 65 nm and a half-bandwidth of 148 nm $\pm$ 20 nm under normal incidence at room temperature. The filter’s peak transmission is about 85\%, and it maintains high transmission blocking for wavelengths beyond 15 µm.

To ensure that the measured detector response originated primarily from the filtered 7.4 µm source rather than unintended thermal radiation, the 0.85 K and 4 K stages of the cryostat were well shielded to prevent leakage of room-temperature blackbody radiation, suppressing stray mid-infrared photons reaching the detector. Consequently, the background photon flux near the antenna resonance wavelength of 7.4 µm was negligible, as evidenced by the absence of any measurable increase in the background count rate of the antenna-coupled SNSPD compared to the bare SNSPD (Figure 2b), despite the enhanced absorption provided by the antenna near 7.4 µm.

The SNSPD output pulse for an antenna-coupled SNSPD at \SI{7.4}{\micro\meter} is shown in Figure \ref{fig:FigS5}b. The multiple rising-edges visible in the output pulse are likely due to reflections caused by impedance mismatch in the readout circuit, and hence the threshold voltage in the counter for PCR and DCR measurements was carefully set to count only the primary pulse while rejecting spurious reflection-induced peaks. In addition, the long reset tail in the output pulse arises from the longer electrical time constants set by the larger series inductance and the 10 $\Omega$ load resistance in the readout circuit. As the maximum PCR of the measured antenna-coupled short-nanowire SNSPD was on the order of $7\times10^{4}$ Hz, this reset tail did not affect the measured PCR. However, for higher PCR devices, such as periodic antenna-coupled nanowire meanders, the readout circuit will need to be optimized to reduce this SNSPD reset tail. Further details of the mid-IR source and SNSPD measurement setup are provided in Taylor \textit{et al.} \cite{taylor2023low}.

\bibliography{mid_antenna_ref}